\documentclass{aa}

\usepackage{graphicx}
\usepackage{natbib}
\usepackage{longtable}
\usepackage{txfonts}
\usepackage{color}
\usepackage[T1]{fontenc}
\usepackage{ulem}

\def\arcsec{\hbox{$^{\prime\prime}$}}

\def\lum{erg s$^{-1}$}

\def\gro{GRO~J1008$-$57}

\newcommand {\be}{\begin {equation}}
\newcommand {\ee}{\end {equation}}
\newcommand {\beq}{\begin {eqnarray}}
\newcommand {\eeq}{\end {eqnarray}}

\usepackage{amstext}

\begin{document}

   \title{Stable accretion from a cold disc in highly magnetized neutron stars}

   \author{S.~S.~Tsygankov \inst{1,2}
          \and A.~A.~Mushtukov \inst{3,2}
          \and V.~F.~Suleimanov \inst{4,2}
          \and V.~Doroshenko \inst{4}
          \and  P.~K.~Abolmasov \inst{1}
          \and  A.~A.~Lutovinov \inst{2,5}
          \and J.~Poutanen \inst{1,6,7}
          }

   \institute{Tuorla Observatory, Department of Physics and Astronomy,  University of Turku,
              V\"ais\"al\"antie 20, FI-21500, Piikki\"o, Finland \\ 
              \email{sergey.tsygankov@utu.fi}
       \and
             Space Research Institute of the Russian Academy of Sciences, Profsoyuznaya Str. 84/32, Moscow 117997, Russia
       \and
            Anton Pannekoek Institute, University of Amsterdam, Science Park 904, 1098 XH Amsterdam, The Netherlands
       \and
            Institut f\"ur Astronomie und Astrophysik, Universit\"at T\"ubingen, Sand 1, D-72076 T\"ubingen, Germany
       \and
             Moscow Institute of Physics and Technology, Moscow region, Dolgoprudnyi, Russia
       \and
          Nordita, KTH Royal Institute of Technology and Stockholm University, Roslagstullsbacken 23, SE-10691 Stockholm, Sweden
       \and 
       Kavli Institute for Theoretical Physics, University of California, Santa Barbara, CA 93106, USA 
          }
   \titlerunning{Accretion from a cold disc in XRPs}
   \authorrunning{S. S. Tsygankov et al. }
   \date{Received; accepted}


  \abstract
  {}
{The aim of this paper is
  to investigate the transition of a strongly magnetized neutron star into the accretion regime
with very low accretion rate. }
  {For this purpose, we monitored the Be-transient X-ray
   pulsar \gro\ throughout a full orbital cycle. The
   current observational campaign was performed with the {\it
     Swift}/XRT telescope in the soft X-ray band (0.5--10 keV) between
   two subsequent Type I outbursts in January and September 2016.}
{The expected transition to the propeller regime was not observed. However, 
transitions between different
   regimes of accretion were detected. In particular, after an outburst the source entered a
   stable accretion state characterised by an accretion rate of
    $\sim10^{14}-10^{15}$ g s$^{-1}$. We associate
   this state with accretion from a cold (low-ionised) disc
   of temperature below $\sim6500$~K. We argue that a transition 
to such accretion regime  should be observed in all X-ray pulsars that have a certain 
combination of the rotation frequency and magnetic field strength. The 
proposed model of  accretion from a cold disc is able to explain 
several puzzling observational properties of X-ray pulsars.
}
   {}


   \maketitle

%

\section{Introduction}

The properties of an accretion disc strongly depend on the parameters of
the compact object and the binary system as a whole. At the same time, they also determine 
the observational appearance of an accreting object. As a consequence, some
of the physical phenomena typical for one class of objects may
not be observable
in another. For instance, transitions between different ionisation states of an
accretion disc are commonly considered to be responsible for bright outbursts 
in dwarf novae and soft X-ray transients \citep[see, e.g. reviews by][]{Ch00,2001NewAR..45..449L,2003cvs..book.....W},
but previously were not considered for X-ray pulsars (XRPs). 
The transition is associated with the so-called thermal-viscous instability that arises as a result of the partial
ionisation of hydrogen in an accretion disc
\citep{1984A&A...132..143M,1984AcA....34..161S,1993PASJ...45..707M,
  1992ApJ...397..664C,1997ApJ...491..312C}.

The ionisation state of the plasma in the disc determines the
disc opacity and the equation
of state, and, therefore, the viscosity (see, e.g. the review by
\citealt{2001NewAR..45..449L}). In a cold disc that consists mainly of atomic
hydrogen, the viscosity and thus the accretion rate are low, so that the accreted matter accumulates in the disc.
In the hot state, the hydrogen is mainly ionised and the disc viscosity dramatically
increases, which allows most of the accumulated matter to be rapidly accreted onto a
compact object. The critical effective
temperature defining the two states is the hydrogen ionisation temperature, which is $\sim 6500$\,K \citep[see Sect.~3.5 in ][]{2003cvs..book.....W}.
The transition between the two states is accompanied by a heating wave
originating either from the inner or outer part
of the accretion disc \citep{1984AcA....34..161S}. The accretion process
becomes stable again when the temperature falls for some reason below the critical value
across the entire disc, that is, in the case of a low mass-accretion rate
\citep{1997ASPC..121..351L},
\be\label{eq:Lasota}
  \dot{M}<\dot M_{\rm cold} \simeq 3.5\times 10^{15}\, r_{10}^{2.65}\,M_{1.4}^{-0.88}~~~{\rm g\,s^{-1}},
\ee
where $r_{10}$ is the inner disc radius\footnote{We define $Q_x=Q/10^x$ in 
cgs units if not stated otherwise.} and $M_{1.4}$ is the neutron star mass in units of
1.4$M_\odot$.

From another perspective, at
very low mass accretion rates onto a magnetized neutron star, the so-called ``propeller effect'' may occur as
an abrupt luminosity drop below some critical value \citep{1975A&A....39..185I}.
This rapid cessation of accretion is caused by a centrifugal barrier produced by
the rotating magnetosphere if it moves faster than the local Keplerian velocity. 
In other words, when the magnetospheric radius 
\be
\label{eq:Rm}
    R_{\rm m} \simeq 2.5\times 10^8\, k \,M_{1.4}^{1/7}\,R_6^{10/7}\,B_{12}^{4/7}\,L_{37}^{-2/7}\,~~~{\rm cm}
\ee
(the neutron star magnetic momentum is taken to be $\mu=B R^3/2$) is larger than the corotation radius, 
\be
     R_{\rm c} = \left(\frac{GMP^2}{4\pi^2} \right)^{1/3} \simeq 1.68 \times 10^8 \,M_{1.4}^{1/3}\,P^{2/3}~~~{\rm cm}.
\ee

The limiting luminosity for the onset of the propeller can be
estimated by equating the corotation and magnetospheric radii \citep[see,
  e.g.][]{2016A&A...593A..16T}:
\be\label{eq_prop} L_{\rm prop}
\simeq \frac{GM\dot{M}_{\rm prop}}{R} \simeq 4 \times 10^{37} k^{7/2}
B_{12}^2 P^{-7/3} M_{1.4}^{-2/3} R_6^5 \,\textrm{erg s$^{-1}$} , 
\ee
where $P$ is the neutron star rotational period in seconds and $\dot{M}$ is the mass-accretion rate.  Factor $k$ accounts for the details
of interaction of the accretion flow with the magnetosphere, relating its size
to the Alfv\'en radius, that is, $k=R_{\rm
  m}/R_{\rm A}$. In the case of disc accretion, it is usually assumed to be $k=0.5$ \citep{GL1978}.

In the case of rapidly rotating accreting XRPs (using 4U~0115+63 and V~0332+53
as case studies), \citet{2016A&A...593A..16T} showed that an
observation of the transition
of the accretion disc to the cold state is impossible because of the onset of the propeller
regime at much higher mass accretion rates than the rate required for the accretion disc to have a
temperature below 6500\,K  at the inner radius. At the
same time, the overall shape and energetics of the outburst were shown
to correspond to expectations from the disc instability model.

XRPs with spin periods that are sufficiently long to reduce the
centrifugal barrier are required to reach a sufficiently low mass-accretion rate
to switch the accretion disc to the cold state before the
transition to the propeller regime.
A few XRPs with long spin periods exist, one of which, \gro,
shows very predictable Type I outbursts every periastron passage with long
quiescence periods between the outbursts \citep{2013A&A...555A..95K}. This source was
selected for the monitoring campaign that covered a full orbital cycle and aimed to
study the source properties at a very low mass-accretion rate.

\gro\ was discovered as an XRP with a period of
$93.587\pm0.005$\,s by the BATSE instrument on board the {\it Compton
  Gamma-Ray Observatory} ({\it CGRO}) during the bright outburst in
1993 \citep{1993IAUC.5836....1S}. Its transient nature was associated with
the Be type of the optical companion (B1-B2\,Ve star;
\citealt{2007MNRAS.378.1427C}). \gro\ shows giant, Type II, and Type
I (associated with the periastron passage) outbursts. The distance to the
source was estimated to be 5.8 kpc \citep{2012A&A...539A.114R}. The
orbital parameters of the binary system were determined using the data
from several observatories: the orbital period $P_{\rm orb}=249.46\pm0.10$ d, the projected
semi-major axis $a_{\rm x}$~sin~$i=530\pm60$~lt-s, the longitude of
periastron $\omega=-26\pm8$~deg, and the eccentricity $e = 0.68\pm0.02$
\citep{2006ATel..940....1L,2007MNRAS.378.1427C, 2012ATel.4564....1K}.
The presence of a cyclotron line at $\sim 88$ keV in the spectrum of the
source was first suggested based on {\it CGRO}/OSSE data
\citep{1999ApJ...512..920S} and was later  confirmed with
\textit{Suzaku} at $E_{\rm cyc} = 75.5$ keV
\citep{2013ATel.4759....1Y}. This allows us to estimate a magnetic
field of the neutron star of $\sim 8\times10^{12}$~G.

In this work we analyse the data obtained with the {\it Swift}/XRT
telescope during one full orbital cycle of Be/XRP \gro\ between the two
consequent Type I outbursts in January and September 2016.  The
unique
combination of high sensitivity and flexibility of scheduling for the
{\it Swift} observatory allowed us to perform a detailed long-term
observational campaign aimed at reaching accretion regimes that
have never been
investigated before. Knowledge of the magnetic field strength of the
neutron star as well as of other parameters of the system allowed us
to interpret the observed temporal behaviour in terms of the
accretion disc instability and to construct the model of accretion
from a cold disc in highly magnetized neutron stars.

\begin{figure}
\centering
\includegraphics[width=0.98\columnwidth, bb=55 155 460 695]{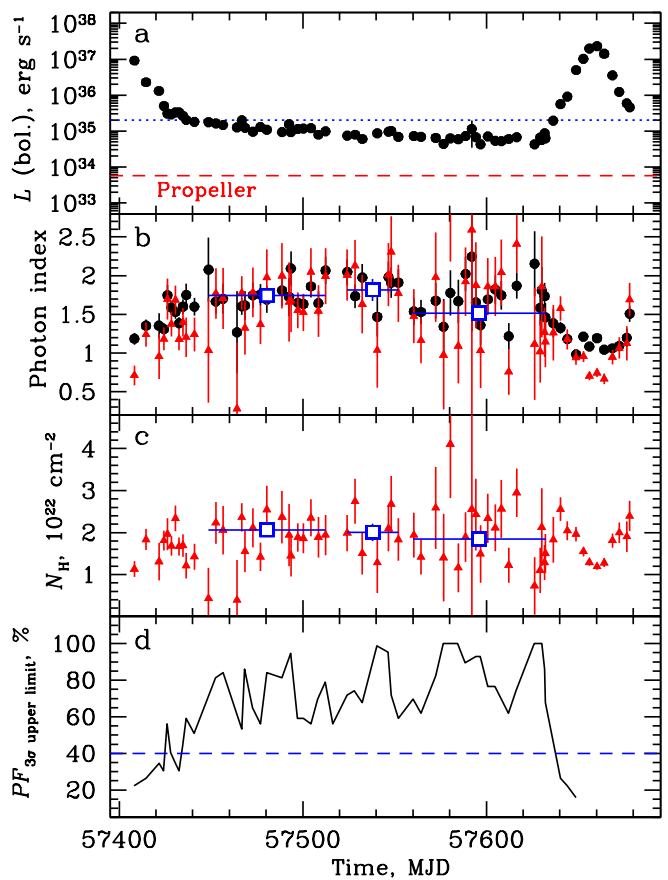}
\caption{(a) Bolometric luminosity of \gro\ obtained by the {\it Swift}/XRT telescope. The luminosity is calculated from the
  unabsorbed flux assuming a distance $d=5.8$ kpc and
  fixed $N_{\rm H}=2.0\times 10^{22}$ cm$^{-2}$.  The dashed red and dotted blue lines
  show the predicted luminosity $L_{\rm prop}$ for the transition to the propeller
  regime and the observed transition to the low-level stable accretion regime, respectively. Panels (b) and (c) represent variations of the photon index and 
absorption value with time, respectively. Black symbols correspond to the fixed $N_{\rm H}$, 
    red triangles and blue open
  squares to the absorption as a free parameter. In panel (d) the estimated $3\sigma$ upper limits for the pulsed fraction as function of time are shown for the {\it Swift}/XRT data with the horizontal dashed line showing the typical pulsed fraction measured during the bright state. The pulsations are clearly detected in all observations, which are expected to be sufficiently sensitive (i.e. below this line), but not elsewhere.} 
  \label{fig:lc}
\end{figure}

\section{Data analysis and results}

\subsection{{\it Swift}/XRT observations}

{\it \textup{The} Swift} observatory \citep{2004ApJ...611.1005G} provides the possibility of
performing long-term observation campaigns of faint X-ray sources using the
onboard focusing XRT telescope \citep{2005SSRv..120..165B}. This work is based
on XRT data collected between two consequent Type I outbursts in January
and September 2016 \citep{2016ATel.8547....1K,2016ATel.9512....1N}.
Observations were performed every 3--4 days in photon-counting (PC) mode. The
spectrum in each observation was extracted using the online
tools\footnote{\url{http://www.swift.ac.uk/user_objects/}} provided by the UK
Swift Science Data Centre \citep[][]{2009MNRAS.397.1177E}.

Spectra were fitted with the {\sc XSPEC} package using the Cash statistic
\citep{1979ApJ...228..939C} after they were grouped to have at least one count
per bin. To avoid any problems caused by the calibration uncertainties at low
energies,\footnote{\url{http://www.swift.ac.uk/analysis/xrt/digest_cal.php}} we
restricted our spectral analysis to the 0.7--10 keV band.

To estimate the bolometric correction factor, we used the results obtained by
\cite{2013A&A...555A..95K}, who demonstrated the dependence of the source
spectrum on its luminosity. Using the spectral parameters in the faint state of
\gro,\ we obtained an estimate for the bolometric correction factor $K_{\rm
bol}\sim1.6$. We note that the remaining uncertainty in the correction factor does
not influence any of the conclusions. In the following analysis, we apply this
correction to the {\it Swift}/XRT data and refer to the bolometric fluxes and
luminosities, unless stated otherwise.

\subsection{Timing analysis}

The light curve of \gro\ obtained from the {\it Swift}/XRT data is shown in
Fig. \ref{fig:lc}. Originally, the observations were requested in order to
detect the transition of \gro\ to the propeller regime (see Section
\ref{sec:discus}) at the limiting luminosity of $L_{\rm prop}=5.6\times10^{33}$
\lum\ (shown in the upper panel of Fig. \ref{fig:lc} with the horizontal dashed
line). However, at some point in the declining phase of the outburst, the source
unexpectedly stopped fading around MJD~57440 at a luminosity
of around $10^{35}$
\lum\ (shown in the same figure with the dotted line), which is one and a half
orders of magnitude higher than $L_{\rm prop}$. During the next seven months,
the source continued to fade, but at a much lower rate characterised by an e-folding
time of $\tau\sim145$ days in contrast to $\tau\sim5.5$ days before the
transition. Strictly speaking, the flux decline during the low-level state is
better fitted with a power-law of index $\sim-0.7$.

Eventually, the monitoring covered the whole orbital cycle, with the sharp
brightening of the source on MJD~57648 signifying the next Type I outburst
during the subsequent periastron passage \citep{2016ATel.9512....1N}. The
transition of the source to the propeller state has therefore not been observed.

\subsubsection{Pulsations}
\label{puls}

Despite the short duration of individual {\it Swift} observations and the long spin
period of the source, the pulsations are easily detectable during the outburst.
Similarly to what has been described in earlier reports \citep{2013A&A...555A..95K,2014ApJ...792..108B}, the pulse profile
exhibits two peaks with a relatively high pulsed fraction of 40$-$50\%, as  is typical for most of XRPs \citep{2009AstL...35..433L}. 
However, no pulsations were detected after the observation
00031030034 (MJD 57430.51), when the source entered the low-level stable accretion state. On the other hand,
the number of photons per observation becomes too low at
this stage (a few tens to a few hundreds), so that the non-detection of the pulsations might simply be due to the insufficient
counting statistics.
For all low-level observations, the $3\sigma$ upper limits for the amplitude
of the pulsations calculated based on the H-test \citep{1989A&A...221..180D} and
following the approach described in \cite{1994MNRAS.268..709B} indeed lie in range
40$-$95\%, as illustrated in Fig.~\ref{fig:lc}d.
The non-detection of the pulsations in individual {\it Swift}
observations at low fluxes is therefore
expected, taking their sensitivity to pulsed flux into account.

On the other hand, a search for the pulsations in the combined light-curve is
complicated by the sparse sampling of the light curve and the
uncertainty in
the orbital parameters of the system. We therefore conclude that the non-detection
of the pulsations by {\it Swift} does not contradict a presence of pulsations given the available counting
statistics and light-curve sampling.

Somewhat stronger limits on the amplitude of the pulsations in the quiescence can
be obtained from the 5\,ks long {\it Chandra} observation 14639 of the
source carried out in May 2013 (MJD 56440.72) between the Type I outbursts. The bolometric
luminosity calculated in the same way as for {\it Swift} data is
$\sim10^{35}$\,\lum, which is comparable with the values measured by {\it Swift}
during our observations in the low-level stable state. To search for the
pulsations, we first extracted photons from a circle with radius 7.4\arcsec\ centred on the
source position after applying the standard screening criteria described in the
instrument documentation. This yields 3215 source photons, which is comparable to the yield for individual {\it Swift}/XRT
observations during the outburst and only a factor of two less than
the total number of photons detected by {\it Swift} in the low-level
state. On the other hand, the {\it Chandra} observation is not affected
by the uncertainty on the orbital parameters. To search for the pulsations, we first corrected the photon arrival times for the effects
of orbital motion in solar and binary systems using the ephemeris reported
by \cite{2013A&A...555A..95K} and then used the
same approach as for the {\it Swift} data. No significant pulsations around the source pulse frequency could be detected with the $3\sigma$ upper limit of $\sim20$\% for periods in
the range 80$-$100\,s.
We note, however, that {\it Chandra} data
are susceptible to pile-up, which potentially reduces the sensitivity to pulsed flux. The pile-up fraction
is lower for pixels in the wings of the point-spread function and higher in its core, meaning that it depends
on the count-rate in a given pixel. Therefore we repeated the simulation above for each pixel within
a circle with radius of 10 pixels centred on the source. All events detected within
a single frame (3.2\,s) were considered as one event and noted to estimate the total pile-up fraction, which is around $\text{}15$\%. The upper limit on the pulsed fraction was then estimated using a concatenated event list
from all pixels as described above, and it indeed was significantly higher at $\sim35$\%.
While slightly lower than the typical pulsed fraction observed during the outburst, 
this by no means excludes pulsations at low fluxes, especially considering the fact that the pulsed fraction
is known to be variable and that we ignored the background for this estimate. We therefore conclude that
additional observations, preferably with {\it XMM-Newton}, are required in order to study the pulsations
in the low-level stable accretion regime.

\subsection{Spectral variability}
\label{sec:specs}

\begin{figure}
\centering
\includegraphics[width=0.98\columnwidth, bb=45 240 550 695]{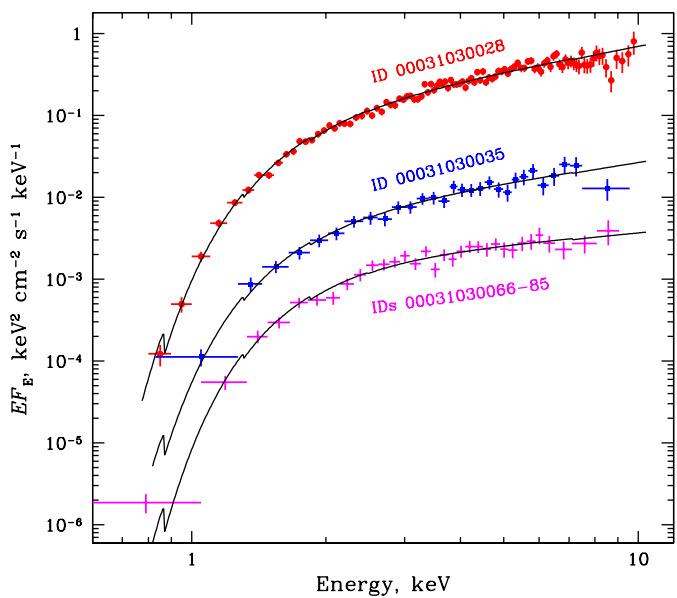}
\caption{Spectra of \gro\ obtained by {\it Swift}/XRT in different
  luminosity states: the brightest state of the January 2016 outburst (ObsId 00031030028; red
  circles), the transiting luminosity (ObsId 00031030035; blue squares),
  and deep inside the stable low-level state (ObsIds 00031030066-85; magenta
  crosses). }\label{fig:spec}
\end{figure}

All spectra from individual observations can be fitted with a simple absorbed
power-law model. To illustrate the stability of the spectral shape over
the luminosity and time, in Fig.~\ref{fig:spec} we present three spectra of
\gro\ obtained in the brightest state of the January 2016 outburst (ObsId
00031030028; red circles), around the transition luminosity (ObsId 00031030035;
blue squares), and deep inside the stable low-level state (ObsIds
00031030066-85; magenta crosses) with luminosities $6.9\times10^{36}$,
$3.1\times10^{35}$ , and $5.7\times10^{34}$ \lum, respectively. No significant
differences are immediately suggested by the data.

A more detailed analysis, however, reveals a minor change in the absorption
and photon index values in the data collected in the bright (outburst)
and low-level states. To obtain sufficient count statistics in the
low-level state (between MJD $57440$ and $57630$), we binned
the available data into three wide bins (see Fig.~\ref{fig:lc} and
Table~\ref{tab1}).

As a result, we found that during the outburst (before
MJD~57440 and after MJD~57648) the mean value of $N_{\rm H}$ is lower than $\sim1.8\times
10^{22}$ cm$^{-2}$ , which is just slightly higher than the Galactic
value in this direction ($N_{\rm H}=1.4\times10^{22}$ cm$^{-2}$;
\citealt{2005A&A...440..775K}). On the other hand, after the transition its value
increases up to $N_{\rm H}=(2.0\pm0.1)\times10^{22}$ cm$^{-2}$. The
difference is most prominent in the parameters of the averaged spectra.
The variability of the spectral parameters during the whole orbital cycle of \gro\ is 
shown in Fig.~\ref{fig:lc} for single and averaged observations.
The middle and bottom panels illustrate variations in photon index and 
absorption column with time, respectively. Black symbols correspond to the absorption column fixed at $N_{\rm
    H}=2.0\times 10^{22}$ cm$^{-2}$, while red triangles and blue open
  squares show the fits with free absorption.

The observational log as well as the best-fit spectral parameters for
the single and averaged observations are presented in Table~\ref{tab1} for the free and fixed 
absorption values.

\section{Discussion}
\label{sec:discus}

Accreting magnetized neutron stars (XRPs, accreting millisecond
pulsars, and accreting magnetars) exhibit a complex behaviour
that is determined by
the interaction of their magnetospheres with the accreting matter. Particularly, 
observational evidence for the propeller effect was found in a number of
sources with magnetic fields from $\sim10^8$ G (for the accreting
millisecond pulsar SAX~J1808.4--3658; \citealt{2008ApJ...684L..99C}) to
$\sim10^{14}$ G (for the first pulsating ULX M82 X-2;
\citealt{2016MNRAS.457.1101T}). The XRPs typically have
magnetic fields somewhere in between these values, and some of
them also exhibit a transition to the propeller regime
\citep{1986ApJ...308..669S,1997ApJ...482L.163C,2001ApJ...561..924C,2016A&A...593A..16T,2017ApJ...834..209L}.

Our observational campaign was originally aimed at detecting the
transition of \gro\ to the propeller regime. In the case of \gro,\
Eq.~(\ref{eq_prop}) gives $L_{\rm prop}=5.6\times10^{33}$ \lum. However,
instead of a sharp drop in flux, \gro\ entered a stable accretion regime
at a luminosity of about $10^{35}$ \lum\ (see Fig.~\ref{fig:lc}), about
one and a half orders of magnitude higher than $L_{\rm prop}$. 

We note that unlike V 0332+53 and 4U 0115+63, where an obvious spectral
softening was observed after the transition to the propeller regime
\citep{2016A&A...593A..16T,2016MNRAS.463L..46W}, no such changes have been observed in
\gro\ after the transition to the stable accretion state. This
strongly suggests that the emission mechanism remains the same,
that is to say, the source continues to accrete even though the data quality was insufficient for detecting pulsations.

\subsection{Accretion disc stability}
\label{sec:instab}

As was shown by \cite{1997ASPC..121..351L}, stable accretion onto the
compact object is possible when the mass accretion rate is high enough
to keep the entire accretion disc hot and fully ionised (temperature
$>6500\,{\rm K}$) or when the mass accretion rate is so low that the
temperature is below the critical value even at the inner radius. At
intermediate levels of the mass accretion rate, the appearance of the
region with a temperature of $6500\,{\rm K}$ will cause a rapidly
propagating heating or cooling front through the disc, leading to the
fast variability of the source flux. In this section we place the
XRPs into the context of the disc instability model and make
predictions for pulsars with different properties \citep[see also
  the discussion in][]{2016A&A...593A..16T}.

We consider  conditions under which an XRP may
accrete matter from the recombined (cold) disc. Particularly,
two criteria have to be fulfilled: (1) the mass accretion rate has to be
high enough to allow the matter to penetrate the centrifugal
barrier, and at the same time, (2) the mass accretion rate has to be sufficiently low
to allow the disc to remain sufficiently cold ($<6500\,{\rm K}$) at every radius.

The first criterion was discussed above and is realized when 
the magnetospheric radius
$R_{\rm m}$ is smaller than the corotation radius $R_{\rm  c}$:
\be \label{com}
     R_{\rm m} \le R_{\rm c}.
\ee
Condition (\ref{com}) is fulfilled  when the accretion luminosity is higher than the
threshold luminosity presented above (see Eq. \ref{eq_prop}): $L \ge L_{\rm prop}$.

The second criterion is satisfied when the mass accretion rate is below the value given by Eq.~(\ref{eq:Lasota})
\citep[$\dot{M}<\dot M_{\rm cold}$; ][]{1997ASPC..121..351L}.
Substituting the magnetospheric radius instead of $r$ into
Eq.~(\ref{eq:Lasota}), we obtain another condition for the stable accretion from
a cold disc:
\be \label{eq:Lcold}
  L < L_{\rm cold}  =  9\times 10^{33}\,k^{1.5}\,M_{1.4}^{0.28}\,R_6^{1.57}\,B_{12}^{0.86}~~~{\rm erg\,s^{-1}}.
\ee
Below this level, the temperature in the accretion disc is lower than 6500\,K at $R>R_{\rm m}$.

If the threshold luminosity for the propeller regime is higher than
the luminosity that corresponds to the transition to the cold disc ($L_{\rm prop} >
L_{\rm cold}$), then the decrease in mass accretion rate during
the outburst decay will cause the transition to the propeller
state. In the opposite case ($L_{\rm prop} <
L_{\rm cold}$), the accretion disc will switch to the cold state with low
viscosity, allowing stable accretion even at a very low rate.

Interestingly, the final state of the source after an outburst is determined by
two fundamental parameters of the neutron star: the magnetic field, and the spin period.
Equating the expressions for luminosities $L_{\rm cold}$ and $L_{\rm prop}$ , we
can derive the critical value of the spin period as a function of the neutron star magnetic
field:
\be \label{eq:Pcrit}
P^{*}=36.6\,k^{6/7}\,B_{12}^{0.49}\,M_{1.4}^{-0.17}\,R_6^{1.22}~~~{\rm s}. 
\ee
If the spin period $P<P^*$, a pulsar will end up in the propeller regime.
Otherwise, the source will start to accrete stably from the cold disc.

\subsection{More accurate condition for accretion from a cold disc}

The condition given by Eq.~(\ref{eq:Lcold}) is based on the assumption that 
the disc temperature is highest at the inner radius given by $R_{\rm
m}$. However, an effective temperature distribution over the radial coordinate
depends on the exact boundary condition and viscous stress at the
magnetospheric radius. Condition (\ref{eq:Lcold}) can be derived more
accurately. There is an uncertainty in the location, where the stress disappears.
This uncertainty is described by the parameter $\beta$: $\beta=1$ corresponds to a
disappearance of the stress at the magnetospheric radius $R_{\rm m}$, while
$\beta =0$ corresponds to the case when the stress disappears at $r \ll R_{\rm
m}$. Then the distribution of effective temperature $T_{\rm eff}$ over the
radial coordinate is given by \citep{2002apa..book.....F}
\be \label{eq:T(r)}
  \sigma_{\rm SB}T_{\rm eff}^4 = \frac{3}{8\pi}\frac{L}{r^3}\,R\,\left[1-\beta \left(\frac{R_{\rm m}}{r}\right)^{1/2}\right],
\ee
where $\sigma_{\rm SB}$ is the Stefan-Boltzmann constant.  
The maximum effective temperature, 
\be
  \sigma_{\rm SB}T_{\rm eff, max}^4 = A\,\frac{3}{8\pi}\frac{L}{R_{\rm m}^3}\,R,
\ee
is achieved at the radius
\be
     r_{\rm max}  = 
     \left\{ \begin{array}{ll}
       \strut\displaystyle 
       \frac{49}{36}\,\beta^2\,R_{\rm m}, & {\rm if}~~~\beta \ge \frac{\sqrt{3}}{2},\\
       R_{\rm m}, & {\rm if}~~~\beta < \frac{\sqrt{3}}{2}, 
       \end{array} \right.
\ee
where
\be
     A  = 
     \left\{ \begin{array}{ll}
       \strut\displaystyle 
       0.057\,\beta^{-6}, & {\rm if}~~~\beta \ge \frac{\sqrt{3}}{2},\\
       1-\beta, & {\rm if}~~~\beta < \frac{\sqrt{3}}{2}.
       \end{array} \right.
\ee
As a result, condition (\ref{eq:Lcold}) can be rewritten in more accurate way. In this approach
the maximum effective temperature in the disc is lower than 
6500\,K throughout the disc when the accretion luminosity is
\be
\label{eq:maxT}
    L \le  L^{(2)}_{\rm cold} \simeq 7\times 10^{33}\!\!A^{-7/13}\,k^{21/13}\,M_{1.4}^{3/13}\,R_6^{23/13} 
      B_{12}^{12/13}\,T_{6500}^{28/13}\ {\rm erg\,s^{-1}},  
\ee
where $T_{6500}=T_{\rm eff}/6500\,{\rm K}$.
This estimate is more accurate
than Eq. (\ref{eq:Lcold}) as it accounts for interaction of the
disc with
the magnetosphere. In the case when the stress disappears at $R_{\rm m}$
, condition (\ref{eq:maxT}) gives $L_{\rm cold}$ higher by a factor of a few
than Eq.~(\ref{eq:Lcold}).

As can be seen from Eq.~\ref{eq:maxT}, our prediction of the transition to the
accretion from the cold disc ($L^{(2)}_{\rm cold}\sim7\times10^{34}$ \lum) coincides
with the observed value within the factor of
2--3. Taking into account existing systematic uncertainties (the
distance to the
system, the physics of the coupling between the disc and the star, the width of the
coupling region, and the uncertainty in the transiting temperature), we consider this
match as support of our physical picture.

Another source of uncertainty can be associated with the irradiation
of the accretion disc by the central object. Some
theoretical models have indeed explained the long duration of the soft X-ray
transients by irradiation of the outer regions of the accretion disc,
which keeps them in hot ionised state for a longer time \citep[see,
  e.g.][]{1998MNRAS.293L..42K}. 
There are also models of the soft X-ray transient bursts that do not
need external irradiation for explanation of accretion discs dynamics
\citep[see, e.g.][and references therein]{2015ApJ...804...87L}, which
might be more appropriate for our case. The observed outburst
duration seems to be compatible with the cooling-wave propagation
timescale. Indeed, the accretion disc is expected to extend to
$4\times 10^{10}$\,cm close to the peak of the outburst (assuming
an effective temperature $T_{\rm eff} = 6500$\,K at this radius).  The
cooling-wave velocity $V_{\rm cw}$ is a few times lower than the
heating wave velocity $V_{\rm hw}$ \citep{1988ApJ...333..227C}, which
can be estimated as $V_{\rm hw} \approx \alpha V_{\rm s}$, where
$V_{\rm s} \approx 10^6$\,cm\,s$^{-1}$ is the sound speed
\citep{1984A&A...131..303M,1993ApJ...419..318C}. The propagation time
of the cooling wave from the radius $4\times 10^{10}$\,cm inward to
$R_{\rm m}$ is therefore expected to be about 10-20 days (assuming
$\alpha=0.1$), which is comparable with observed duration.

\begin{figure}
\centering
\includegraphics[width=0.98\columnwidth, bb=55 270 565 675]{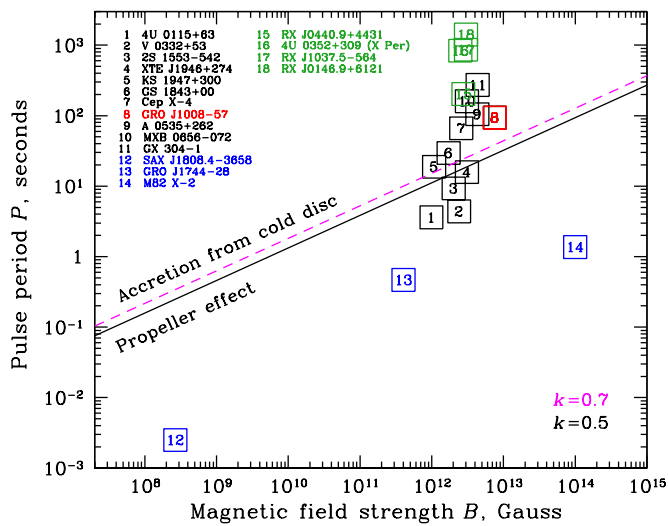}
\caption{Collection of some known BeXRPs (shown in black), as
  well as the accreting millisecond  pulsar SAX J1808.4$-$3658, the 
  intermediate pulsar GRO J1744$-$28, and the accreting magnetar M82 X-2
  (all three shown in blue) on the $B-P$ plane. The solid  and
  dashed lines correspond to the prediction of $P^{*}(B)$ from
  Eq.~(\ref{eq:Pcrit2}) for $\beta=1$ and two different values of parameter $k$: 0.5
  and 0.7, respectively. This line separates sources where the
propeller regime is possible from sources with stable accretion from the cold disc. Persistent low-luminous BeXRPs are shown in green. \gro\ is shown in red and resides in the area corresponding
  to the sources with accretion from the cold disc. 
  }\label{fig:prop}
\end{figure}

Using the updated Eq.~(\ref{eq:maxT}) for the critical accretion
luminosity $L^{(2)}_{\rm cold}$ , we can derive an updated equation for
the critical value of the spin period $P^{*}$ , which, as described
above, determines the pulsar behaviour at very low mass-accretion
rate. Now instead of Eq.~(\ref{eq:Pcrit}) we have
\be \label{eq:Pcrit2}
P^{*}=40.7\,k^{21/23}\,B_{12}^{6/13}\,M_{1.4}^{-5/13}\,R_6^{18/13}\,A^{3/13}\,T_{6500}^{-12/13}~~~{\rm
  s}.  \ee This equation can be used to predict the behaviour of any
XRP (with a known magnetic field) expected in the case of transient
activity.  This is especially important for XRPs in Be binary systems
(BeXRPs), which are known to be transient sources \citep[see,
  e.g.][]{2011Ap&SS.332....1R} with magnetic fields measured from the
cyclotron lines in their spectra \citep{2015A&ARv..23....2W}. For
illustration, we show some known BeXRPs as well as the accreting
millisecond pulsar SAX J1808.4$-$3658, the intermediate pulsar
GRO~J1744$-$28, and the accreting magnetar M82 X-2 on the $B-P$ plane
in Fig.~\ref{fig:prop}. All sources in this plane are divided with the
prediction from Eq.~(\ref{eq:Pcrit2}) into two groups: (i) those
entering the propeller regime at a low mass-accretion rate (below the
line), and (ii) sources where stable accretion from the cold disc
continues at any accretion rate (above the line). Persistent
low-luminous BeXRPs \citep{1999MNRAS.306..100R} are shown in
green. Clearly, \gro\ resides in the area corresponding to the
sources with accretion from the cold disc. The majority of sources
residing below the $P^{*}(B)$ line have previously been shown to exhibit
transitions to the propeller regime \citep[for a review
  see][]{2016A&A...593A..16T}. It is important to note that peak
  luminosities in BeXRPs outbursts are ranging from $10^{37}$ to
  $10^{39}$~\lum, exceeding all possible values of $L_{\rm cold}$ and
  $L_{\rm prop}$. During the outburst decline, such sources therefore
  inevitably end up in one of the above-mentioned states.

\section{Conclusion}

We analysed the {\it Swift}/XRT observations of
\gro\ obtained between two subsequent Type I outbursts in January and
September 2016. The original idea was to detect the transition of the
source to the propeller regime, which is accompanied by an abrupt decrease in source flux and a softening of its spectrum. However, during the
declining phase of the outburst, the source unexpectedly stopped its
fading and entered a stable accretion state that was characterised by an
accretion rate of the order of $\sim10^{14}-10^{15}$ g s$^{-1}$ and a
hard spectrum.  We associate this state with accretion from the
cold
(low-ionised) disc with a temperature below $\sim6500$~K.

We proposed a model of accretion from the cold disc in the systems harbouring
neutron stars with strong magnetic fields (i.e. XRPs). The basic idea of the model is that in slowly rotating
neutron stars the centrifugal barrier caused by the rotating
magnetosphere is greatly suppressed in comparison to the fast rotating
neutron stars, leading to  a much lower threshold luminosity for the transition
to the propeller regime. This allows such sources to reach mass-accretion rates that are so low that the temperature throughout the accretion disc
becomes lower than the hydrogen recombination limit of $\sim 6500\,{\rm
  K}$.

When this mass accretion rate is reached, the fast fading of the source
intensity should stop and further accretion with a low rate in the stable
accretion regime from the cold disc with very low viscosity is
expected to continue.
This behaviour was observed in the pulsar \gro\ between two consequent Type I
outbursts in January and September 2016.

Our model has strong predictive power. Particularly, the transition to the
accretion regime from a cold disc is expected to be observed in all XRPs with
a certain combination of pulse period and magnetic field strength.
Other manifistations of the cold disc accretion associated with a
change in disc structure could be anticipated.  A change of the inner
disc radius can be expected to affect the spin evolution of the pulsar,
aperiodic variability properties, pulse profiles, and the energy spectrum
of the source. A detailed calculation of the accretion disc structure in
this case is ongoing and will be published elsewhere.

\begin{acknowledgements}

We are grateful to Ilia Potravnov for a number of useful comments and to the Swift team for the execution of our ToO
request.  This work was supported by the Russian Science Foundation grant 14-12-01287
(SST, AAM, AAL, VS), 
the Academy of Finland grant 268740, and the National Science Foundation grant PHY-1125915 (JP). 
VD  thanks the Deutsches Zentrums for Luft- und Raumfahrt (DLR)
and the DFG for financial support (grant DLR~50~OR~0702). 
We also acknowledge the support of COST Action MP1304.
\end{acknowledgements}

\bibliographystyle{aa}
\bibliography{allbib}

\begin{thebibliography}{47}
\expandafter\ifx\csname natexlab\endcsname\relax\def\natexlab#1{#1}\fi

\bibitem[{{Bellm} {et~al.}(2014){Bellm}, {F{\"u}rst}, {Pottschmidt}, {Tomsick},
  {Boggs}, {Chakrabarty}, {Christensen}, {Craig}, {Hailey}, {Harrison},
  {Stern}, {Walton}, {Wilms}, \& {Zhang}}]{2014ApJ...792..108B}
{Bellm}, E.~C., {F{\"u}rst}, F., {Pottschmidt}, K., {et~al.} 2014, \apj, 792,
  108

\bibitem[{{Brazier}(1994)}]{1994MNRAS.268..709B}
{Brazier}, K.~T.~S. 1994, \mnras, 268, 709

\bibitem[{{Burrows} {et~al.}(2005){Burrows}, {Hill}, {Nousek}, {Kennea},
  {Wells}, {Osborne}, {Abbey}, {Beardmore}, {Mukerjee}, {Short}, {Chincarini},
  {Campana}, {Citterio}, {Moretti}, {Pagani}, {Tagliaferri}, {Giommi},
  {Capalbi}, {Tamburelli}, {Angelini}, {Cusumano}, {Br{\"a}uninger}, {Burkert},
  \& {Hartner}}]{2005SSRv..120..165B}
{Burrows}, D.~N., {Hill}, J.~E., {Nousek}, J.~A., {et~al.} 2005, \ssr, 120, 165

\bibitem[{{Campana} {et~al.}(2001){Campana}, {Gastaldello}, {Stella}, {Israel},
  {Colpi}, {Pizzolato}, {Orlandini}, \& {Dal Fiume}}]{2001ApJ...561..924C}
{Campana}, S., {Gastaldello}, F., {Stella}, L., {et~al.} 2001, \apj, 561, 924

\bibitem[{{Campana} {et~al.}(2008){Campana}, {Stella}, \&
  {Kennea}}]{2008ApJ...684L..99C}
{Campana}, S., {Stella}, L., \& {Kennea}, J.~A. 2008, \apjl, 684, L99

\bibitem[{{Cannizzo}(1993)}]{1993ApJ...419..318C}
{Cannizzo}, J.~K. 1993, \apj, 419, 318

\bibitem[{{Cannizzo} {et~al.}(1988){Cannizzo}, {Shafter}, \&
  {Wheeler}}]{1988ApJ...333..227C}
{Cannizzo}, J.~K., {Shafter}, A.~W., \& {Wheeler}, J.~C. 1988, \apj, 333, 227

\bibitem[{{Cash}(1979)}]{1979ApJ...228..939C}
{Cash}, W. 1979, \apj, 228, 939

\bibitem[{{Chen} {et~al.}(1997){Chen}, {Shrader}, \&
  {Livio}}]{1997ApJ...491..312C}
{Chen}, W., {Shrader}, C.~R., \& {Livio}, M. 1997, \apj, 491, 312

\bibitem[{{Cheng} {et~al.}(1992){Cheng}, {Horne}, {Panagia}, {Shrader},
  {Gilmozzi}, {Paresce}, \& {Lund}}]{1992ApJ...397..664C}
{Cheng}, F.~H., {Horne}, K., {Panagia}, N., {et~al.} 1992, \apj, 397, 664

\bibitem[{{Cherepashchuk}(2000)}]{Ch00}
{Cherepashchuk}, A.~M. 2000, \ssr, 93, 473

\bibitem[{{Coe} {et~al.}(2007){Coe}, {Bird}, {Hill}, {McBride}, {Schurch},
  {Galache}, {Wilson}, {Finger}, {Buckley}, \&
  {Romero-Colmenero}}]{2007MNRAS.378.1427C}
{Coe}, M.~J., {Bird}, A.~J., {Hill}, A.~B., {et~al.} 2007, \mnras, 378, 1427

\bibitem[{{Cui}(1997)}]{1997ApJ...482L.163C}
{Cui}, W. 1997, \apjl, 482, L163

\bibitem[{{de Jager} {et~al.}(1989){de Jager}, {Raubenheimer}, \&
  {Swanepoel}}]{1989A&A...221..180D}
{de Jager}, O.~C., {Raubenheimer}, B.~C., \& {Swanepoel}, J.~W.~H. 1989, \aap,
  221, 180

\bibitem[{{Evans} {et~al.}(2009){Evans}, {Beardmore}, {Page}, {Osborne},
  {O'Brien}, {Willingale}, {Starling}, {Burrows}, {Godet}, {Vetere}, {Racusin},
  {Goad}, {Wiersema}, {Angelini}, {Capalbi}, {Chincarini}, {Gehrels}, {Kennea},
  {Margutti}, {Morris}, {Mountford}, {Pagani}, {Perri}, {Romano}, \&
  {Tanvir}}]{2009MNRAS.397.1177E}
{Evans}, P.~A., {Beardmore}, A.~P., {Page}, K.~L., {et~al.} 2009, \mnras, 397,
  1177

\bibitem[{{Frank} {et~al.}(2002){Frank}, {King}, \&
  {Raine}}]{2002apa..book.....F}
{Frank}, J., {King}, A., \& {Raine}, D.~J. 2002, {Accretion Power in
  Astrophysics: Third Edition} (Cambridge, UK: Cambridge University Press), 398

\bibitem[{{Gehrels} {et~al.}(2004){Gehrels}, {Chincarini}, {Giommi}, {Mason},
  {Nousek}, {Wells}, {White}, {Barthelmy}, {Burrows}, {Cominsky}, {Hurley},
  {Marshall}, {M{\'e}sz{\'a}ros}, {Roming}, {Angelini}, {Barbier}, {Belloni},
  {Campana}, {Caraveo}, {Chester}, {Citterio}, {Cline}, {Cropper}, {Cummings},
  {Dean}, {Feigelson}, {Fenimore}, {Frail}, {Fruchter}, {Garmire}, {Gendreau},
  {Ghisellini}, {Greiner}, {Hill}, {Hunsberger}, {Krimm}, {Kulkarni}, {Kumar},
  {Lebrun}, {Lloyd-Ronning}, {Markwardt}, {Mattson}, {Mushotzky}, {Norris},
  {Osborne}, {Paczynski}, {Palmer}, {Park}, {Parsons}, {Paul}, {Rees},
  {Reynolds}, {Rhoads}, {Sasseen}, {Schaefer}, {Short}, {Smale}, {Smith},
  {Stella}, {Tagliaferri}, {Takahashi}, {Tashiro}, {Townsley}, {Tueller},
  {Turner}, {Vietri}, {Voges}, {Ward}, {Willingale}, {Zerbi}, \&
  {Zhang}}]{2004ApJ...611.1005G}
{Gehrels}, N., {Chincarini}, G., {Giommi}, P., {et~al.} 2004, \apj, 611, 1005

\bibitem[{{Ghosh} \& {Lamb}(1978)}]{GL1978}
{Ghosh}, P. \& {Lamb}, F.~K. 1978, \apjl, 223, L83

\bibitem[{{Illarionov} \& {Sunyaev}(1975)}]{1975A&A....39..185I}
{Illarionov}, A.~F. \& {Sunyaev}, R.~A. 1975, \aap, 39, 185

\bibitem[{{Kalberla} {et~al.}(2005){Kalberla}, {Burton}, {Hartmann}, {Arnal},
  {Bajaja}, {Morras}, \& {P{\"o}ppel}}]{2005A&A...440..775K}
{Kalberla}, P.~M.~W., {Burton}, W.~B., {Hartmann}, D., {et~al.} 2005, \aap,
  440, 775

\bibitem[{{King} \& {Ritter}(1998)}]{1998MNRAS.293L..42K}
{King}, A.~R. \& {Ritter}, H. 1998, \mnras, 293, L42

\bibitem[{{Kretschmar} {et~al.}(2016){Kretschmar}, {Kuehnel}, {Nakajima},
  {Rothschild}, {Laplace}, {Salazar}, {Mihara}, {Kreykenbohm}, \&
  {Wilson-Hodge}}]{2016ATel.8547....1K}
{Kretschmar}, P., {Kuehnel}, M., {Nakajima}, M., {et~al.} 2016, The
  Astronomer's Telegram, 8547

\bibitem[{{Kuehnel} {et~al.}(2012){Kuehnel}, {Mueller}, {Kreykenbohm}, {Wilms},
  {Pottschmidt}, {Fuerst}, {Rothschild}, {Caballero}, {Klochkov}, {Staubert},
  {Suchy}, {Kretschmar}, {Ferrigno}, {Torrejon}, \&
  {Martinez-Nunez}}]{2012ATel.4564....1K}
{Kuehnel}, M., {Mueller}, S., {Kreykenbohm}, I., {et~al.} 2012, The
  Astronomer's Telegram, 4564

\bibitem[{{K{\"u}hnel} {et~al.}(2013){K{\"u}hnel}, {M{\"u}ller}, {Kreykenbohm},
  {F{\"u}rst}, {Pottschmidt}, {Rothschild}, {Caballero}, {Grinberg},
  {Sch{\"o}nherr}, {Shrader}, {Klochkov}, {Staubert}, {Ferrigno},
  {Torrej{\'o}n}, {Mart{\'{\i}}nez-N{\'u}{\~n}ez}, \&
  {Wilms}}]{2013A&A...555A..95K}
{K{\"u}hnel}, M., {M{\"u}ller}, S., {Kreykenbohm}, I., {et~al.} 2013, \aap,
  555, A95

\bibitem[{{Lasota}(1997)}]{1997ASPC..121..351L}
{Lasota}, J.~P. 1997, in Astronomical Society of the Pacific Conference Series,
  Vol. 121, IAU Colloq. 163: Accretion Phenomena and Related Outflows, ed.
  D.~T. {Wickramasinghe}, G.~V. {Bicknell}, \& L.~{Ferrario}, 351

\bibitem[{{Lasota}(2001)}]{2001NewAR..45..449L}
{Lasota}, J.-P. 2001, \nar, 45, 449

\bibitem[{{Levine} \& {Corbet}(2006)}]{2006ATel..940....1L}
{Levine}, A.~M. \& {Corbet}, R. 2006, The Astronomer's Telegram, 940

\bibitem[{{Lipunova}(2015)}]{2015ApJ...804...87L}
{Lipunova}, G.~V. 2015, \apj, 804, 87

\bibitem[{{Lutovinov} \& {Tsygankov}(2009)}]{2009AstL...35..433L}
{Lutovinov}, A.~A. \& {Tsygankov}, S.~S. 2009, Astronomy Letters, 35, 433

\bibitem[{{Lutovinov} {et~al.}(2017){Lutovinov}, {Tsygankov}, {Krivonos},
  {Molkov}, \& {Poutanen}}]{2017ApJ...834..209L}
{Lutovinov}, A.~A., {Tsygankov}, S.~S., {Krivonos}, R.~A., {Molkov}, S.~V., \&
  {Poutanen}, J. 2017, \apj, 834, 209

\bibitem[{{Meyer}(1984)}]{1984A&A...131..303M}
{Meyer}, F. 1984, \aap, 131, 303

\bibitem[{{Meyer} \& {Meyer-Hofmeister}(1984)}]{1984A&A...132..143M}
{Meyer}, F. \& {Meyer-Hofmeister}, E. 1984, \aap, 132, 143

\bibitem[{{Mineshige} {et~al.}(1993){Mineshige}, {Yamasaki}, \&
  {Ishizaka}}]{1993PASJ...45..707M}
{Mineshige}, S., {Yamasaki}, T., \& {Ishizaka}, C. 1993, \pasj, 45, 707

\bibitem[{{Nakajima} {et~al.}(2016){Nakajima}, {Kawase}, {Negoro}, {Mihara},
  {Tomida}, {Ueno}, {Nakahira}, {Ishikawa}, {Sugawara}, {Nakagawa}, {Sugizaki},
  {Serino}, {Iwakiri}, {Shidatsu}, {Sugimoto}, {Takagi}, {Matsuoka}, {Kawai},
  {Isobe}, {Sugita}, {Yoshii}, {Tachibana}, {Ono}, {Fujiwara}, {Harita},
  {Muraki}, {Yoshida}, {Sakamoto}, {Kawakubo}, {Kitaoka}, {Tsunemi}, {Shomura},
  {Tanaka}, {Masumitsu}, {Ueda}, {Kawamuro}, {Hori}, {Tanimoto}, {Tsuboi},
  {Nakamura}, {Sasaki}, {Yamauchi}, {Furuya}, \&
  {Yamaoka}}]{2016ATel.9512....1N}
{Nakajima}, M., {Kawase}, T., {Negoro}, H., {et~al.} 2016, The Astronomer's
  Telegram, 9512

\bibitem[{{Reig}(2011)}]{2011Ap&SS.332....1R}
{Reig}, P. 2011, \apss, 332, 1

\bibitem[{{Reig} \& {Roche}(1999)}]{1999MNRAS.306..100R}
{Reig}, P. \& {Roche}, P. 1999, \mnras, 306, 100

\bibitem[{{Riquelme} {et~al.}(2012){Riquelme}, {Torrej{\'o}n}, \&
  {Negueruela}}]{2012A&A...539A.114R}
{Riquelme}, M.~S., {Torrej{\'o}n}, J.~M., \& {Negueruela}, I. 2012, \aap, 539,
  A114

\bibitem[{{Shrader} {et~al.}(1999){Shrader}, {Sutaria}, {Singh}, \&
  {Macomb}}]{1999ApJ...512..920S}
{Shrader}, C.~R., {Sutaria}, F.~K., {Singh}, K.~P., \& {Macomb}, D.~J. 1999,
  \apj, 512, 920

\bibitem[{{Smak}(1984)}]{1984AcA....34..161S}
{Smak}, J. 1984, \actaa, 34, 161

\bibitem[{{Stella} {et~al.}(1986){Stella}, {White}, \&
  {Rosner}}]{1986ApJ...308..669S}
{Stella}, L., {White}, N.~E., \& {Rosner}, R. 1986, \apj, 308, 669

\bibitem[{{Stollberg} {et~al.}(1993){Stollberg}, {Finger}, {Wilson}, {Harmon},
  {Rubin}, {Zhang}, \& {Fishman}}]{1993IAUC.5836....1S}
{Stollberg}, M.~T., {Finger}, M.~H., {Wilson}, R.~B., {et~al.} 1993, \iaucirc,
  5836

\bibitem[{{Tsygankov} {et~al.}(2016{\natexlab{a}}){Tsygankov}, {Lutovinov},
  {Doroshenko}, {Mushtukov}, {Suleimanov}, \& {Poutanen}}]{2016A&A...593A..16T}
{Tsygankov}, S.~S., {Lutovinov}, A.~A., {Doroshenko}, V., {et~al.}
  2016{\natexlab{a}}, \aap, 593, A16

\bibitem[{{Tsygankov} {et~al.}(2016{\natexlab{b}}){Tsygankov}, {Mushtukov},
  {Suleimanov}, \& {Poutanen}}]{2016MNRAS.457.1101T}
{Tsygankov}, S.~S., {Mushtukov}, A.~A., {Suleimanov}, V.~F., \& {Poutanen}, J.
  2016{\natexlab{b}}, \mnras, 457, 1101

\bibitem[{{Walter} {et~al.}(2015){Walter}, {Lutovinov}, {Bozzo}, \&
  {Tsygankov}}]{2015A&ARv..23....2W}
{Walter}, R., {Lutovinov}, A.~A., {Bozzo}, E., \& {Tsygankov}, S.~S. 2015,
  \aapr, 23, 2

\bibitem[{{Warner}(2003)}]{2003cvs..book.....W}
{Warner}, B. 2003, {Cataclysmic Variable Stars} (Cambridge: Cambridge
  University Press), 592

\bibitem[{{Wijnands} \& {Degenaar}(2016)}]{2016MNRAS.463L..46W}
{Wijnands}, R. \& {Degenaar}, N. 2016, \mnras, 463, L46

\bibitem[{{Yamamoto} {et~al.}(2013){Yamamoto}, {Mihara}, {Sugizaki}, {Sasano},
  {Makishima}, \& {Nakajima}}]{2013ATel.4759....1Y}
{Yamamoto}, T., {Mihara}, T., {Sugizaki}, M., {et~al.} 2013, The Astronomer's
  Telegram, 4759, 1

\end{thebibliography}

\newpage


\onecolumn

\begin{longtable}{ccccccc}
\caption[{\it Swift}/XRT observations of the source \gro]{{\it Swift}/XRT observations of \gro.} \label{tab1}\\ 
\hline\hline
Obs Id & Date &  Exposure & \multicolumn{2}{c}{$N_{\rm H}=2.0\times10^{22}$~cm$^{-2}$} & \multicolumn{2}{c}{Free $N_{\rm H}$} \\
\cline{4-5}
\cline{6-7}
       &  MJD  &  ks        & XRT flux$^{a}$                 & $\Gamma$$^{b}$                   & $N_{\rm H}$         & $\Gamma$$^{b}$             \\
       &       &            & $10^{-11}$ erg s$^{-1}$ cm$^{-2}$ &                     & 10$^{22}$ cm$^{-2}$  &              \\
\hline
00031030028 & 57408.1683 & 1.0 & 142$\pm$5 & 1.2$\pm$0.1 & 1.1$\pm$0.2  & 0.7$\pm$0.1  \\
00031030029 & 57414.4910 & 1.3 & 35.5$\pm$1.3 & 1.4$\pm$0.1 & 1.8$\pm$0.3  & 1.2$\pm$0.2  \\
00031030030 & 57421.6127 & 1.2 & 20.3$\pm$1.5 & 1.4$\pm$0.2 & 1.3$\pm$0.4  & 1.0$\pm$0.3  \\
00031030031 & 57424.0809 & 2.5 & 7.8$\pm$0.4 & 1.3$\pm$0.1 & 1.8$\pm$0.2  & 1.2$\pm$0.1  \\
00031030032 & 57426.0158 & 1.2 & 4.8$^{+0.3}_{-0.2}$ & 1.7$\pm$0.1 & 2.0$\pm$0.4  & 1.7$\pm$0.3  \\
00031030033 & 57428.0596 & 2.3 & 4.6$^{+0.2}_{-0.1}$ & 1.6$\pm$0.1 & 1.7$\pm$0.2  & 1.4$\pm$0.2  \\
00031030034 & 57430.5115 & 2.7 & 5.1$^{+0.2}_{-0.1}$ & 1.5$\pm$0.1 & 2.3$\pm$0.3  & 1.7$\pm$0.2  \\
00031030035 & 57432.5746 & 2.9 & 5.1$^{+0.2}_{-0.1}$ & 1.4$\pm$0.1 & 1.7$\pm$0.2  & 1.2$\pm$0.1  \\
00031030036 & 57434.5145 & 2.5 & 4.1$\pm$0.2 & 1.6$\pm$0.1 & 1.7$\pm$0.3  & 1.4$\pm$0.2  \\
00031030037 & 57436.4258 & 1.3 & 3.2$\pm$0.2 & 1.7$\pm$0.1 & 1.2$\pm$0.3  & 1.2$\pm$0.3  \\
00031030038 & 57440.8239 & 2.3 & 2.8$\pm$0.1 & 1.6$\pm$0.1 & 1.4$\pm$0.3  & 1.2$\pm$0.2  \\
00031030040 & 57448.6581 & 0.2 & 2.8$^{+0.8}_{-0.5}$ & 2.1$\pm$0.4 & 0.4$\pm$0.7  & 1.0$\pm$0.7  \\
00031030041 & 57452.4376 & 1.3 & 2.5$\pm$0.2 & 1.7$\pm$0.2 & 2.2$\pm$0.5  & 1.8$\pm$0.3  \\
00031030042 & 57456.4917 & 1.2 & 2.3$\pm$0.2 & 1.7$\pm$0.2 & 2.1$\pm$0.6  & 1.7$\pm$0.4  \\
00031030043 & 57464.1391 & 0.2 & 2.0$^{+0.5}_{-0.4}$ & 1.3$\pm$0.5 & 0.4$\pm$1.0  & 0.3$\pm$0.9  \\
00031030044 & 57466.6732 & 2.4 & 3.1$^{+0.2}_{-0.1}$ & 1.6$\pm$0.1 & 2.4$\pm$0.3  & 1.8$\pm$0.2  \\
00031030045 & 57468.2726 & 1.3 & 1.9$^{+0.1}_{-0.2}$ & 1.6$\pm$0.2 & 1.6$\pm$0.5  & 1.3$\pm$0.3  \\
00031030046 & 57472.5946 & 2.4 & 1.5$\pm$0.1 & 1.7$\pm$0.2 & 2.1$\pm$0.5  & 1.8$\pm$0.3  \\
00031030047 & 57476.8448 & 2.6 & 2.0$\pm$0.1 & 1.8$\pm$0.1 & 1.4$\pm$0.3  & 1.4$\pm$0.3  \\
00031030048 & 57480.1727 & 1.8 & 1.7$^{+0.2}_{-0.1}$ & 1.7$\pm$0.2 & 2.5$\pm$0.6  & 2.0$\pm$0.4  \\
00031030049 & 57488.6718 & 1.3 & 1.5$\pm$0.1 & 1.8$\pm$0.2 & 2.4$\pm$0.6  & 2.0$\pm$0.4  \\
00031030050 & 57492.4004 & 0.6 & 2.4$^{+0.2}_{-0.3}$ & 1.7$\pm$0.2 & 1.9$\pm$0.7  & 1.7$\pm$0.5  \\
00031030051 & 57493.4566 & 1.3 & 1.5$\pm$0.2 & 2.1$\pm$0.2 & 1.4$\pm$0.5  & 1.7$\pm$0.4  \\
00031030052 & 57497.0396 & 3.1 & 1.8$\pm$0.1 & 1.6$\pm$0.1 & 1.9$\pm$0.3  & 1.6$\pm$0.2  \\
00031030053 & 57500.2303 & 2.9 & 1.8$\pm$0.1 & 1.6$\pm$0.1 & 1.9$\pm$0.4  & 1.5$\pm$0.2  \\
00031030054 & 57504.3630 & 2.2 & 1.9$\pm$0.1 & 1.9$\pm$0.1 & 2.4$\pm$0.4  & 2.0$\pm$0.3  \\
00031030055 & 57508.1415 & 2.9 & 1.2$\pm$0.1 & 1.6$\pm$0.2 & 1.9$\pm$0.5  & 1.5$\pm$0.3  \\
00031030056 & 57512.2662 & 1.9 & 1.5$^{+0.2}_{-0.1}$ & 2.1$\pm$0.2 & 1.9$\pm$0.5  & 2.0$\pm$0.4  \\
00031030058 & 57524.0342 & 2.8 & 1.2$\pm$0.1 & 2.0$\pm$0.2 & 2.0$\pm$0.4  & 2.0$\pm$0.3  \\
00031030059 & 57528.0203 & 2.9 & 1.2$\pm$0.1 & 1.7$\pm$0.2 & 2.7$\pm$0.5  & 2.1$\pm$0.3  \\
00031030060 & 57532.4048 & 2.9 & 0.9$\pm$0.1 & 2.0$\pm$0.2 & 1.5$\pm$0.5  & 1.6$\pm$0.4  \\
00031030061 & 57540.3100 & 1.1 & 1.4$^{+0.1}_{-0.2}$ & 1.5$\pm$0.3 & 1.3$\pm$0.7  & 1.0$\pm$0.5  \\
00031030062 & 57546.4969 & 1.5 & 1.5$\pm$0.1 & 2.0$\pm$0.2 & 2.1$\pm$0.5  & 2.0$\pm$0.4  \\
00031030063 & 57548.0885 & 1.6 & 1.6$^{+0.2}_{-0.1}$ & 1.9$\pm$0.2 & 2.7$\pm$0.7  & 2.3$\pm$0.5  \\
00031030064 & 57552.0065 & 2.9 & 1.1$\pm$0.1 & 1.9$\pm$0.2 & 1.8$\pm$0.5  & 1.8$\pm$0.4  \\
00031030066 & 57560.0381 & 2.6 & 1.1$\pm$0.1 & 1.5$\pm$0.2 & 1.9$\pm$0.5  & 1.5$\pm$0.3  \\
00031030067 & 57564.2988 & 3.0 & 1.1$\pm$0.1 & 1.5$\pm$0.2 & 1.4$\pm$0.4  & 1.2$\pm$0.3  \\
00031030068 & 57572.2871 & 1.7 & 1.0$\pm$0.1 & 1.7$\pm$0.2 & 2.6$\pm$1.0  & 2.0$\pm$0.6  \\
00031030069 & 57576.4712 & 1.6 & 0.7$\pm$0.1 & 1.3$\pm$0.4 & 1.4$\pm$1.0  & 1.0$\pm$0.7  \\
00031030070 & 57580.3262 & 1.0 & 1.0$\pm$0.1 & 1.8$\pm$0.3 & 4.1$\pm$1.3  & 2.9$\pm$0.7  \\
00031030071 & 57584.3187 & 1.3 & 0.9$\pm$0.1 & 1.7$\pm$0.3 & 1.2$\pm$0.6  & 1.1$\pm$0.5  \\
00031030072 & 57588.4948 & 1.6 & 1.1$\pm$0.1 & 2.0$\pm$0.2 & 1.9$\pm$0.6  & 1.9$\pm$0.4  \\
00031030073 & 57592.0136 & 0.1 & 1.8$^{+1.8}_{-0.7}$ & 2.2$\pm$1.0 & 2.6$\pm$3.2  & 2.6$\pm$2.4  \\
00031030074 & 57594.2706 & 1.4 & 1.1$\pm$0.1 & 1.7$\pm$0.3 & 2.4$\pm$0.9  & 1.9$\pm$0.6  \\
00031030075 & 57596.7272 & 2.7 & 0.66$^{+0.08}_{-0.07}$ & 1.4$\pm$0.3 & 1.5$\pm$0.7  & 1.0$\pm$0.5  \\
00031030076 & 57600.7122 & 2.6 & 1.1$\pm$0.1 & 1.7$\pm$0.2 & 2.3$\pm$0.6  & 1.9$\pm$0.4  \\
00031030077 & 57604.6330 & 3.7 & 0.83$^{+0.08}_{-0.07}$ & 1.8$\pm$0.2 & 2.1$\pm$0.6  & 1.9$\pm$0.4  \\
00031030078 & 57608.0913 & 3.3 & 0.81$^{+0.08}_{-0.07}$ & 1.7$\pm$0.2 & 2.6$\pm$0.7  & 2.0$\pm$0.4  \\
00031030079 & 57612.0746 & 3.4 & 0.93$^{+0.07}_{-0.08}$ & 1.2$\pm$0.2 & 1.2$\pm$0.4  & 0.8$\pm$0.3  \\
00031030080 & 57616.2037 & 3.1 & 1.1$\pm$0.1 & 1.9$\pm$0.2 & 3.0$\pm$0.6  & 2.4$\pm$0.4  \\
00031030081 & 57626.1855 & 1.4 & 0.7$^{+0.2}_{-0.1}$ & 2.2$\pm$0.4 & 0.7$\pm$0.7  & 1.1$\pm$0.7  \\
00031030082 & 57629.0364 & 1.7 & 1.1$\pm$0.1 & 1.6$\pm$0.2 & 1.1$\pm$0.5  & 1.0$\pm$0.4  \\
00031030083 & 57630.1009 & 1.1 & 0.9$\pm$0.1 & 1.8$\pm$0.3 & 2.1$\pm$0.9  & 1.9$\pm$0.7  \\
00031030084 & 57631.6256 & 2.1 & 1.4$\pm$0.1 & 1.7$\pm$0.2 & 1.3$\pm$0.4  & 1.3$\pm$0.3  \\
00031030085 & 57632.0994 & 2.8 & 0.98$^{+0.07}_{-0.09}$ & 1.5$\pm$0.2 & 1.5$\pm$0.5  & 1.1$\pm$0.3  \\
00031030086 & 57636.4050 & 0.8 & 3.0$^{+0.2}_{-0.3}$ & 1.4$\pm$0.2 & 1.8$\pm$0.5  & 1.3$\pm$0.4  \\
00031030087 & 57640.3242 & 3.2 & 8.7$^{+0.4}_{-0.2}$ & 1.3$\pm$0.1 & 2.6$\pm$0.3  & 1.6$\pm$0.2  \\
00031030088 & 57644.0456 & 3.4 & 14.1$^{+0.5}_{-0.4}$ & 1.2$\pm$0.1 & 2.0$\pm$0.2  & 1.2$\pm$0.1  \\
00031030089 & 57648.6950 & 3.6 & 78$\pm$2 & 1.0$\pm$0.1 & 2.0$\pm$0.2  & 0.9$\pm$0.1  \\
00031030090 & 57652.7537 & 3.3 & 159$\pm$3 & 1.2$\pm$0.1 & 1.5$\pm$0.1  & 1.0$\pm$0.1  \\
00031030091 & 57656.0031 & 3.3 & 308$^{+4}_{-5}$ & 1.1$\pm$0.1 & 1.3$\pm$0.1  & 0.7$\pm$0.1  \\
00031030092 & 57660.1913 & 3.5 & 360$^{+6}_{-5}$ & 1.2$\pm$0.1 & 1.2$\pm$0.1  & 0.7$\pm$0.1  \\
00031030093 & 57664.1124 & 3.1 & 219$\pm$4 & 1.0$\pm$0.1 & 1.3$\pm$0.1  & 0.7$\pm$0.1  \\
00031030094 & 57668.5570 & 3.4 & 54.9$\pm$1.5 & 1.1$\pm$0.1 & 1.8$\pm$0.2  & 0.9$\pm$0.1  \\
00031030095 & 57672.0878 & 2.7 & 19.1$^{+0.6}_{-0.7}$ & 1.1$\pm$0.1 & 2.0$\pm$0.2  & 1.1$\pm$0.1  \\
00031030096 & 57676.0145 & 1.2 & 9.1$^{+0.4}_{-0.6}$ & 1.2$\pm$0.1 & 1.9$\pm$0.4  & 1.1$\pm$0.2  \\
00031030097 & 57678.0084 & 1.6 & 7.1$\pm$0.1 & 1.5$\pm$0.1 & 2.4$\pm$0.4  & 1.7$\pm$0.2  \\
\multicolumn{7}{c}{Averaged points}\\
00031030040-56 & -- & 29.7 & 1.74$\pm$0.04 & 1.74$\pm$0.06 & 2.06$\pm$0.08  & 1.75$\pm$0.09  \\
00031030058-64 & -- & 15.6 & 1.15$\pm$0.05 & 1.84$\pm$0.07 & 2.0$\pm$0.2  & 1.8$\pm$0.2  \\
00031030066-85 & -- & 42.2 & 0.95$\pm$0.03 & 1.63$\pm$0.05 & 1.9$\pm$0.2  & 1.5$\pm$0.1  \\
\hline



\end{longtable}
\begin{center}
$^{a}$   Unabsorbed flux in the 0.5--10 keV energy range derived from the model with a fixed absorption value. \\
$^{b}$  Photon index.   
\end{center}


\twocolumn

\end{document}